# Atomic-scale images of charge ordering in a mixed-valence manganite.

Ch. Renner[*], G. Aeppli[*], B-G. Kim[†], Yeong-Ah Soh[*] and S.-W. Cheong[†]

[*]*NEC Research Institute, 4, Independence Way, Princeton, NJ 08540, USA*

[†]*Rutgers University, Department of Physics and Astronomy, Piscataway, NJ 08854, USA*

**Transition-metal perovskite oxides exhibit a wide range of extraordinary but imperfectly understood phenomena. Charge, spin, orbital, and lattice degrees of freedom all undergo order-disorder transitions in regimes not far from where the best-known of these phenomena, namely high-temperature superconductivity of the copper oxides[1], and the 'colossal' magnetoresistance of the manganese oxides[2,3], occur. Mostly diffraction techniques, sensitive either to the spin or the ionic core, have been used to measure the order. Unfortunately, because they are only weakly sensitive to valence electrons and yield superposition of signals from distinct mesoscopic phases, they cannot directly image mesoscopic phase coexistence and charge ordering, two key features of the manganites. Here we describe the first experiment to image charge ordering and phase separation in real space with atomic-scale resolution in a transition metal oxide. Our scanning tunneling microscopy (STM) data show that charge order is correlated with structural order, as well as with whether the material is locally metallic or insulating, thus giving an atomic-scale basis for descriptions[4] of the manganites as mixtures of electronically and structurally distinct phases.**

The material chosen for our experiments is $Bi_{1-x}Ca_xMnO_3$ (BCMO). For trivalent Bi and divalent Ca, the Mn ions are in a mixed valence state $Mn^{3+x}$. At high temperature, $Mn^{3+}$ and $Mn^{4+}$ randomly occupy the manganese sites. Upon reducing the temperature, these cations are believed to order, yielding an increased lattice periodicity visible to X-ray and neutron diffraction[5]. For our nominally x=0.76 samples, grown from a BiO flux, this occurs at $T_{CO}$=250K, as established using SQUID magnetometry. We performed the STM experiments in ultrahigh vacuum at a base pressure of 5 $10^{-10}$ Torr. Previously published STM investigations of manganites primarily focused on spectroscopy of the density of states averaged over many atoms[6,7], and demonstrated phase separation into metallic and insulating regions on submicron[8], but not atomic length scales. In contrast, STM with atomic resolution has already been achieved for cuprates and revealed inhomogeneities in the superconducting order on atomic length scales[9,10].

BCMO single crystals do not cleave naturally, and preparing flat and atomically clean surfaces suitable for STM is a major challenge. It turned out that cleaning the as-grown samples in ethanol with a cotton stick shortly before loading into the vacuum chamber, allowed us to obtain reproducible topographic images with atomic resolution. The STM tips were made of etched tungsten wires. Typical tunneling parameters during imaging were 0.2nA and 0.7V for the set current and sample bias respectively. The piezoelectric scan actuators were calibrated to an accuracy of 4% by imaging layered graphite crystals at different temperatures.

As one might expect based on the rather crude surface preparation, STM failed to yield atomic resolution over large portions of the surface. We mainly saw a broad range of different nanometer scale textures. However, we repeatedly found clean terraces a few 100nm$^2$ wide where atomic resolution was routinely achieved. The room temperature image in Fig. 1A clearly shows the square lattice expected for the (010) face of BCMO (simple cubic notation). The corresponding lattice constant $a_0=3.8\pm0.6$Å is in good agreement with the value of 3.77Å determined by X-ray diffraction[11,12]. At room temperature, we sometimes also observe a '$\sqrt{2}a_0 \times \sqrt{2}a_0$' lattice, with the unit cell doubled along [101], coexisting with the ordinary square lattice (Fig. 2A). The cut, shown in Fig. 2B, along one crystallographic axis in Fig. 2A suggests a homogeneous charge distribution among the atomic sites in the cubic (disordered) region, whereas charges appear redistributed among alternating atomic sites in the ordered region, generating the doubled unit cell. The most natural interpretation is that there is local charge ordering, namely the alternation of $Mn^{3+}$ and $Mn^{4+}$ ions, as posited in the simplest descriptions of mixed valence manganites. Conductance spectra acquired in all areas where the doubled unit cell is stable systematically show more insulating characteristics than those measured in cubic regions (Fig. 2C). This validates the association of charge ordering with a metal-insulator transition – the carriers are localized and the material is insulating when charges are distributed in a set spatial pattern rather than fluctuating with time at each site. The metallic state is not an especially good metal, with a semi-metallic spectrum superposed on a small ohmic term (inset to Fig. 2C). The energy at which the deviations from ohmic behavior become apparent is around 0.1 eV, which is also where a strong non-Drude contribution begins to dominate the optical conductivity for similar samples at room temperature[13]. The gap for the insulating spectrum collected for the regions at room temperature where the doubled unit cell occurs, is ~0.7 eV. This is very similar to the gap one can deduce from the optical conductivity measured at 150K, when the bulk of the crystal is charge-ordered[13]. At a qualitative level, the spectra of Fig. 2C look strikingly similar to those measured by Fäth *et al.*[8] on the isostructural compound $(La,Ca)MnO_3$. While their lower hole doped samples lack a charge-ordering phase transition, their finding of spectroscopic inhomogeneities near the metal-insulator transition is very consistent with our results. Fig. 2 is remarkable because it shows the first data to connect the different tunneling spectra to distinct phases at the atomic scale.

When the sample is cooled below $T_{CO}$, the $\sqrt{2}a_0 \times \sqrt{2}a_0$ lattice becomes the dominant structure (Fig. 3). The current-voltage spectra of these regions always show insulating characteristics strikingly similar to those measured in the ordered regions above $T_{CO}$, as illustrated in Fig 4. Fig. 3A displays a large region of uninterrupted $\sqrt{2}a_0 \times \sqrt{2}a_0$ order of precisely the type seen at the upper right of Fig. 2A. There is a clear modulation of the height as well as the positions of the atoms away from the vertices of a square lattice. A more thorough analysis of Fig. 3A reveals a finer structure, characterized by alternating long and short interatomic distances (ID) along the main crystallographic axes (Fig. 3B). The short ID form regular zigzag chains across the entire scanned area (yellow lines in Figs. 3A and 3D), thus breaking the crystal symmetry. The distortion amplitude, which is the difference between long and short ID, can be inferred directly from histograms of all ID measured along the cubic unit cell vectors (Fig. 3C). This distortion, generally in excess of 0.5Å, is too large to ascribe to the Mn atoms alone. On the other hand, the bright spots in STM images such as Fig. 3A, match the apical oxygens sitting over the manganese sites on the (010) surface according to X-ray experiments[11] (Fig. 3D). The X-ray data[11] show that the $MnO_6$ octahedra are tilted by an angle $\theta$ of about 10°, hence shifting the apical oxygen by a distance $d_{//}=0.34$Å away from the vertices of the square lattice, where $d_{//}=a_{MO}\sin(\theta)$, and $a_{MO}=1.94$Å is the Mn-O bond length. The tilting is not uniform, rather the octahedra form a zigzag pattern where neighbouring oxygens are either separated or brought closer by roughly $2d_{//}$. For a lattice constant of 3.77Å, these shifts should yield a lattice with alternating interatomic distances of 3.1Å and 4.4Å, in excellent agreement with our STM images (Fig.3C). Interestingly, the room temperature images bear a similar distribution of short and long ID. However, in contrast to the low temperature images, the short ID are more randomly distributed (Fig. 1B). Static disorder in the oxygen tilts[14,15] therefore appears to be *annealed* on cooling through $T_{CO}$, yielding a regular zigzag pattern. Although the precise temperature where this happens at the surface has not been identified in the present experiments, these results illustrate on a local scale the idea that ordered lattice distortions, especially tilts of the $MnO_6$ octahedra, are important for stabilizing the charge-ordered state in manganites[16].

Fig. 3A contains not only information about the geometry of the crystal face exposed, but also about the 'heights' of its features. It is natural to ask whether the apparent 'height' modulation is derived simply from a modulation of atomic coordinates (possible surface reconstruction), or whether it indicates a modulation of the electronic wavefunctions centered at the atom cores. The last paragraph's analysis of the in-plane displacements shows that the STM contrast consistently reflects the tilting of the $MnO_6$ octahedra. The apex oxygens illustrated as the larger blue spheres in Fig. 3D, move a certain distance perpendicular to the (010) plane when the octahedra distort and tilt due to the Jahn-Teller (JT) effect induced by a central $Mn^{3+}$. Since all octahedra are tilted by approximately the same angle ($\theta \approx 10°$), the apex oxygens are all displaced by roughly the same distance $d_{\perp}=a_{MO}(1-\cos(\theta))$. Thus, the tilting of the octahedra cannot account for the observed corrugation. On the other hand, the JT distortion of the order 0.05-0.1Å[11,17] happens exclusively for octahedra centered at $Mn^{3+}$. Taking this together with the fact that the octahedra are tilted, the JT distortion will then account for at most 0.1Å of the height modulation at the apical oxygen sites. What we see instead – as clearly demonstrated by the histogram in Fig. 3E – are apparent height differences between neighbouring atomic sites which are frequently larger than 0.1 Å. Thus, the measured corrugation needs an additional contribution, which is most likely due to different wavefunctions responsible for the tunneling between the two neighbouring octahedra and the STM tip. The most natural explanation is that the Mn ions at the centers of these octahedra have different charges and therefore different orbital occupancy, which affects the tunneling through the apical oxygens that cap the planes exposed to our STM experiments.

The charge-ordered STM images discussed so far consistently show a $Mn^{3+}$ to $Mn^{4+}$ ratio of 1:1, in apparent contradiction with diffraction data[5,12] (which show a periodicity of ~$4\sqrt{2}a_0$ rather than the $\sqrt{2}a_0$ we see) for crystals prepared following the same procedures and with the same nominal doping. There are many possible explanations for this discrepancy. The first is to appeal to the surface sensitivity of STM, and assert that the surfaces behave differently from the bulk. Indeed, such an argument has been made for $Sr_2RuO_4$, the only other transition metal oxide without copper for which atomic resolution STM data have been reported[18]. The agreement of our atomic coordinates with those deduced from bulk X-ray crystallography (Fig. 3D) speaks against a surface reconstruction on the scale of that seen for $Sr_2RuO_4$. Furthermore, the agreement of the tunneling gap shown in Fig. 4 with the optical gap[13] (another bulk probe) also speaks against a surface effect. This leaves the possibility that even though the bulk charge ordering transition for our crystals is as sharp as any previously reported, the effective doping may not be x=0.76 at all surfaces. Besides the surface, which may affect the actual local doping, there is the possibility of phase separation into (commensurate) charge-ordered phases upon cooling through $T_{CO}$[19]. It is worth noting that what distinguishes our experiments from previous STM work on manganites is that we have atomic resolution for a small, rather than zero fraction of the crystal surface exposed. This leaves the possibility that the remaining fraction would exhibit structures consistent with bulk diffraction data for the nominal composition of the sample. The fact that the measured optical gap[13] corresponds to a smooth rise in conductivity rather than a hard onset is consistent with coexistence, even in the bulk, of distinct ordered phases with different periodicities and different gaps.

We have presented the first atomic-scale images of any manganese oxide using scanning tunneling microscopy. The images display many of the phenomena that have been posited for the manganites, most notably charge ordering, of which STM is the most direct probe because of its sensitivity to the outer valence electrons – diffraction and transmission electron microscopy, which have been the tools for the initial exploration of the charge ordering phenomena, are all primarily sensitive to atomic core positions rather than the outer electrons. We have been able to associate the metallic and insulating current-voltage characteristics with distinct atomic-scale structures.

1. Orenstein, J. & Millis, A. J. Advances in the physics of high-temperature superconductivity. *Science* **288**, 468-474 (2000).

2. von Helmolt, R., Wecker, J., Holzapfel, B., Schultz, L. & Samwer, K. Giant negative magnetoresistance in perovskitelike $La_{2/3}Ba_{1/3}MnO_x$ ferromagnetic films. *Phys. Rev. Lett.* **71**, 2331-2333 (1993).

3. Jin, S. *et al*. Thousand–fold change in resistivity in magnetoresistive La-Ca-Mn-O films. *Science* **264**, 413-415 (1994).

4. Moreo, A., Yunoki, S. & Dagotto, E. Phase separation scenario for manganese oxides and related materials. *Science* **283**, 2034-2040 (1999).

5. Bao, W., Axe, J. D., Chen, C. H. & Cheong, S-W. Impact of charge ordering on magnetic correlations in perovskite (Bi,Ca)$MnO_3$. *Phys. Rev. Lett.* **78**, 543-546 (1997).

6. Wei, J. Y. T., Yeh, N.-C. & Vasquez, R. P. Tunneling evidence of half-metallic ferromagnetism in $La_{0.7}Ca_{0.3}MnO_3$. *Phys. Rev. Lett*. **79**, 5150-5153 (1997).

7. Biswas, A., Elizabeth, S., Raychaudhuri, A. K. & Bhat, H. L. Density of states of hole-doped manganites: A scanning-tunneling-microscopy/spectroscopy study. *Phys. Rev. B* **59**, 5368-5376 (1999).

8. Fäth, M. *et al.* Spatially inhomogeneous metal-insulator transition in doped manganites. *Science* **285**, 1540-1542 (1999).

9. Howald, C., Fournier, P. & Kapitulnik A. Inherent inhomogeneities in tunneling spectra of $Bi_2Sr_2CaCu_2O_{8-x}$ crystals in the superconducting state. *Phys. Rev. B* **64**, 100504(R) 1-4 (2001).

10. Pan, S.H. *et al.* Microscopic electronic inhomogeneity in the high-$T_c$ superconductor $Bi_2Sr_2CaCu_2O_{8+x}$. *Nature* **413**, 282-285 (2001).

11. Radaelli, P. G., Cox, D. E., Marezio, M. & Cheong, S-W. Charge, orbital, and magnetic ordering in $La_{0.5}Ca_{0.5}MnO_3$. *Phys. Rev. B* **55**, 3015-3023 (1997).

12. Su, Y., Du, C.-H., Hatton, P. D., Collins, S. P. & Cheong S-W. Charge ordering and the related structural phase transition in single-crystal $(Bi_{0.24}Ca_{0.76})MnO_3$. *Phys. Rev. B* **59**, 11687-11692 (1999).

13. Liu, H. L., Cooper, S. L. & Cheong, S-W. Optical study of the evolution of charge and spin ordering in the manganese perovskite $Bi_{1-x}Ca_xMnO_3$ (x>0.5). *Phys. Rev. Lett.* **81,** 4684-4687 (1998).

14. Billinge, S. J. L., DiFrancesco, R. G., Kwei, G. H., Neumeier, J. J. & Thomson, J. D. Direct observation of lattice polaron formation in the local structure of $La_{1-x}Ca_xMnO3$. *Phys. Rev. Lett.* **77**, 715-718 (1996).

15. Millis, A. J. Lattice effects in magnetoresistive perovskites. *Nature* **392**, 147-150 (1998).

16. Woo, H., Tyson, T. A., Croft, M., Cheong, S-W. & Woicik, J. C. Correlation between the magnetic and structural properties of Ca-doped $BiMnO_3$. *Phys. Rev. B* **63**, 134412 1-12 (2001).

17. Radaelli, P. G., Cox. D. E., Capogna, L., Cheong, S-W. & Marezio, M. Wigner-crystal and bi-stripe models for the magnetic and crystallographic superstructures of $La_{0.333}Ca_{0.667}MnO_3$, *Phys. Rev. B* **59**, 14440-14450 (1999).

18. Matzdorf, R. *et al*. Ferromagnetism Stabilized by Lattice Distortion at the Surface of the p-Wave Superconductor $Sr_2RuO_4$. *Science* **289**, 746-748 (2000).

19. Kagan, M. Yu., Kugel, K. I. & Khomskii, D. I. Phase separation in systems with charge ordering, *J. Exp. Theor. Phys*. **93**, 415-423 (2001).

**Acknowledgements**. We are grateful to N. Wingreen and J. Chadi for comments on the manuscript, and acknowledge helpful discussions with D. Khomskii, A . Millis, C. de Morais Smith, and A. Yazdani. BGK and SWC are supported by the National Science Foundation.

Correspondence should be addressed to C.R. (e-mail: renner@research.nj.nec.com).

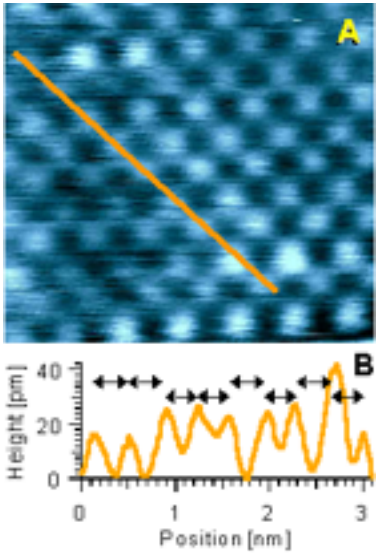

**Figure 1** STM mapping of the paramagnetic phase of $Bi_{0.24}Ca_{0.76}MnO_3$ at 299 Kelvin. **A** 3.4x3.1 nm$^2$ image with well resolved square lattice ($a_0$=3.8±0.6Å). **B** The intensity profile reveals a random distribution of short and long interatomic distances (indicated by arrows) along the main crystallographic axes.

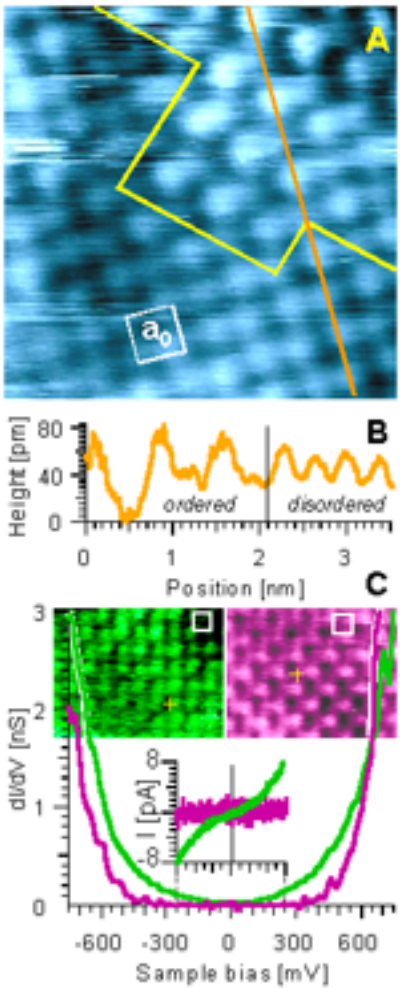

**Figure 2** Topographic and spectroscopic atomic scale signatures of phase separation into metallic and insulating regions in the paramagnetic phase of $Bi_{0.24}Ca_{0.76}MnO_3$ at 299 K. **A** 3.5x3.5 nm$^2$ STM image of a grain boundary (yellow line) between an insulating $\sqrt{2}a_0 \times \sqrt{2}a_0$ charge-ordered region (upper right) and a more metallic homogeneous cubic region (lower left). **B** Intensity profile extracted along the orange line in A. Note the larger amplitude modulation in the ordered region due to charge ordering. **C** Charge-ordered regions with the $\sqrt{2}a_0 \times \sqrt{2}a_0$ lattice (purple) yield insulating dI/dV(V) characteristics, while the disordered cubic regions (green) are characterized by more metallic dI/dV(V) characteristics (numerical derivatives normalized to the metallic junction resistance R=V/I at 0.7V). The low bias part of the corresponding I(V) data are shown in the inset. The spectra were taken at the yellow crosses on the 3.7x2.9 nm$^2$ STM images (white squares = cubic unit cell).

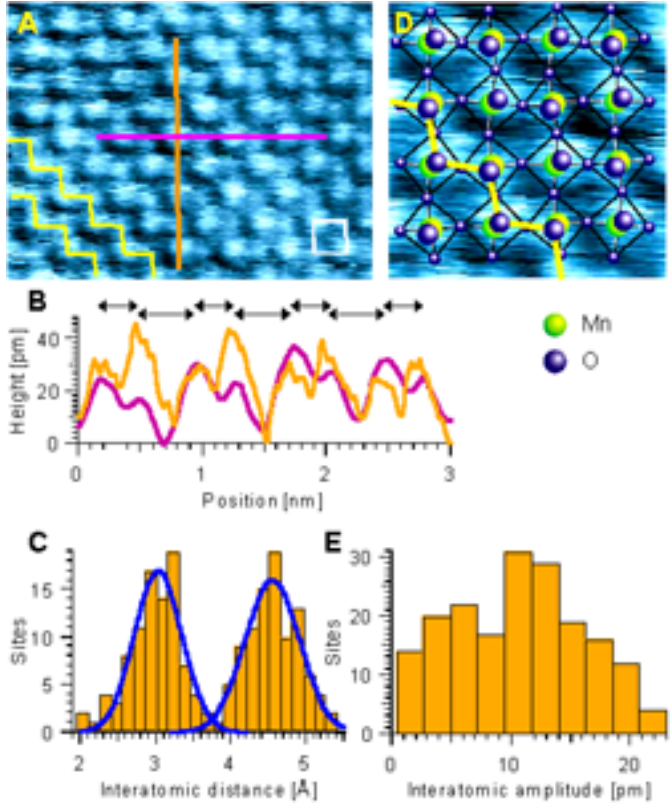

**Figure 3** Atomic scale STM mapping of the charge ordered phase of $Bi_{0.24}Ca_{0.76}MnO_3$ at 146 Kelvin. **A** 4.8x3.6 nm$^2$ image of the $\sqrt{2}a_0 \times \sqrt{2}a_0$ lattice (white square = cubic unit cell). **B** Intensity profiles extracted along the main crystallographic directions in A. They show the distortion of the atomic positions away from the cubic vertices with alternating short and long interatomic distances (arrows). **C** Bimodal distribution into short (3.0±0.2Å) and long (4.5±0.3Å) interatomic distances along [100] and [001]. **D** X-ray refined crystal structure of the (010) plane of the isostructural compound $La_{1-x}Ca_xMnO_3$ (x=½) adapted from Ref.11 superposed on a magnified area of A (1.7x1.7 nm$^2$). For purposes of illustration, the apical oxygens are sketched larger than the in-plane oxygen. **E** The amplitude difference between neighbouring atomic sites peaks above 10pm, too large to be solely due to structural distortions determined by X-ray diffraction.

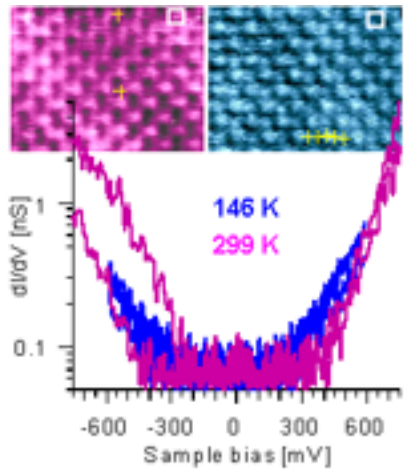

**Figure 4** Identical lattice and electronic structures of the insulating $\sqrt{2}a_0 \times \sqrt{2}a_0$ regions observed in the paramagnetic room temperature phase and the charge-ordered low temperature phase. The gap in the differential tunneling conductance (numerical derivatives normalized to the 299 Kelvin junction resistance R=V/I at 0.8V) is essentially the same (~0.7eV) at 299 Kelvin (purple) and at 146 Kelvin (blue). The STM constant current images (4.5x3.5 nm$^2$) reveal the same lattice structures in regions where the insulating spectra were measured (yellow crosses). The white squares correspond to the cubic unit cell. For the 8 (out of 34) W tips for which atomic resolution was obtained, there was 100% correspondence between the different structural phases and the conducting or insulating nature of the spectra.